\documentclass[12pt]{article}
\usepackage{amsmath,amssymb,mathtools,amsfonts,epsfig,graphicx,euscript}%,mathrsfs}
 \usepackage{color}
 \usepackage{etoolbox}
\newcommand{\nn}{\nonumber\\}

\begin{document}
\begin{titlepage}
\thispagestyle{empty}

\vspace{4cm}
\begin{center}
\font\titlerm=cmr10 scaled\magstep4
\font\titlei=cmmi10
scaled\magstep4 \font\titleis=cmmi7 scaled\magstep4 {\Large{\textbf{The cosmic QCD phase transition with dense matter and its gravitational waves from holography}
\\}}
\setcounter{footnote}{0}
\vspace{1.5cm} \noindent{{%\large
M. Ahmadvand \footnote{e-mail:ahmadvand@shahroodut.ac.ir} and K. Bitaghsir
Fadafan \footnote{e-mail:bitaghsir@shahroodut.ac.ir
}
}}\\
\vspace{0.2cm}

{\it Faculty of Physics, Shahrood University of Technology, P.O.Box 3619995161 Shahrood, Iran\\}

\vspace*{.4cm}

\end{center}
\vskip 2em
\setcounter{footnote}{0}
%----------------------------------------------------------------------------
\begin{abstract}
Consistent with cosmological constraints, there are scenarios with the large lepton asymmetry which can lead to the finite baryochemical potential at the cosmic QCD phase  transition scale. In this paper, we investigate this possibility in the holographic models. Using the holographic renormalization method, we find the first order Hawking-Page phase transition, between Reissner-Nordstr$\rm\ddot{o}$m AdS black hole and thermal charged AdS space, corresponding to the de/confinement phase transition. We obtain the gravitational wave spectra generated during the evolution of bubbles for a range of the bubble wall velocity and examine the reliability of the scenarios and consequent calculations by gravitational wave experiments.

\end{abstract}
\end{titlepage}
%%%%%%%%%%%%%%%%%%%%%%%%%%%%
\section{Introduction}%%%%%%
%%%%%%%%%%%%%%%%%%%%%%%%%%%%

Cosmological Phase Transitions (PTs) in the early universe have played significant roles in the universe that we can see at the present time. The last PT at which quarks and gluons are confined is the phase transition of Quantum Chromodynamics (QCD) which occurred around $ 10^{-5} $ secs after the big bang. The theory of QCD describes strong nuclear interactions. Despite other theories in the standard model of particle physics, QCD is strongly-coupled at low energy and cannot be described by perturbative methods, though at high energy it is an asymptotic freedom theory. For light quarks, the QCD Lagrangian has an approximate symmetry called chiral symmetry, which is spontaneously broken, in the QCD vacuum. Chiral condensate as the order parameter of this PT becomes non-zero in the broken phase, and pseudo-Goldstone pions and conservation of the baryon number are remainders of this spontaneously broken symmetry. Furthermore, there is another approximate symmetry, the global Z(3) center symmetry which is spontaneously broken under the de/confinement PT, for heavy quarks. For this PT, the expectation value of Polyakov loop is the relevant order parameter, which can be obtained from the heavy quark potential \cite{Fukushima:2010bq, Cai:2012xh}.\\
If a PT is first order, it gives rise to non-equilibrium events, the nucleation and growth of bubbles. Two degenerate states with minimum free energy are separated by the bubbles. The vacuum energy of the system causes bubbles to expand and collide with each other. The spherical symmetry of the bubbles is broken and parts of the energy can create Gravitational Waves (GWs) in the spacetime \cite{Kosowsky:1992rz}. The process of the bubble collision can also produce bulk motion which itself is taken into account as another source for GWs through sound waves \cite{Hindmarsh:2013xza} and Magnetohydrodynamic (MHD) turbulence \cite{Caprini:2006jb} in the plasma. Due to the weakness of gravitational interactions and least attenuation of GWs, the detection of their signal gives us important information from early epochs of the universe.\\
At the energies which the two mentioned aspects of the QCD PT occur, the theory is strongly-coupled, thus perturbative expansions cannot be applied. Lattice QCD as a numerical method can help to understand these phenomena. In this approach, it is shown that for 2+1 intermediate bare quark masses (2 light quarks, up and down, and 1 heavier quark, strange) with negligible baryochemical potential, PT is not first order but a crossover \cite{Aoki:2006br}, whereas for so heavy, static quarks or pure gauge theory, PT is first order \cite{Lucini:2012wq}. However, for the finite baryochemical potential, this method suffers from the sign problem related to the complexity of fermion determinant \cite{Takeda:2011vd}.\\
Here, we focus on the de/confinement aspect of the PT at the finite baryochemical potential. The baryochemical potential depends on baryon and lepton asymmetries and for tiny baryon and lepton asymmetries it would vanish. However, a large lepton asymmetry can be supported in the early universe \cite{Schwarz:2009ii}. This finite baryochemical potential can be justified in late leptogenesis scenarios \cite{Hambye:2001eu}, which are compatible with BBN and CMB constraints. The large lepton asymmetry is expected to be in the neutrino chemical potential \cite{Schwarz:2009ii} and can be proposed in the models with the dark matter neutrino candidate \cite{Boyarsky:2009ix}.\\
In this work, we use AdS/QCD approach to explaining the de/confinement PT. The conjecture of gauge/gravity as an extension to AdS/CFT correspondence has been a useful technique describing a strongly-coupled gauge theory by a gravitational theory in a higher dimensional spacetime \cite{Ramallo:2013bua}. Deriving universal properties of these theories, such as the ratio of the shear viscosity of the hot plasma to the entropy density, is one of the issues handled through these dualities \cite{Policastro:2001yc}. People also tried to address QCD PT features within this context. In \cite{Witten:1998zw}, it is shown that there is a correspondence between the first order Hawking-Page (H-P) PT and the de/confinement PT for compact boundaries. Also, for non-compact boundaries with the removed small radius region of AdS space, \cite{Herzog:2006ra} found the H-P PT.\\
The study of GWs from the cosmological QCD PT using the gauge/gravity duality was initiated in \cite{Ahmadvand:2017xrw}. In that paper, we studied the cosmological QCD PT considering gluodynamics and zero baryochemical potential. We used AdS/QCD models to find the corresponded H-P PT and calculated the GW spectra radiated during the PT. In this paper, we are interested in studying holographically the cosmological deconfinement transition with possible finite baryochemical potential. Taking quark degrees of freedom into account leads to adding an abelian gauge field on the gravity side. (For the holographic QCD by considering finite chemical potential see \cite{Horigome:2006xu, Lee:2009bya, Park:2011qq}.) Within hard and soft wall models \cite{Polchinski:2001tt, Karch:2006pv} we here apply the holographic renormalization \cite{Emparan:1999pm} to find the H-P PT, between Reissner-Nordstr$\rm\ddot{o}$m AdS black hole (RN AdS BH) and thermal charged AdS (tc AdS).\\ 
For zero baryochemical potential, the temperature at the PT is determined by a special horizon radius which is fixed by IR cut-off in the models \cite{Herzog:2006ra}. However, as we will see, in the case of finite baryochemical potential, temperature depends on baryochemical potential as well. Therefore, to specify temperature and baryochemical potential at the transition, we also investigate the string configuration based on the expectation value of Polyakov loop as the order parameter during the PT. Finally, We extend our approach in \cite{Ahmadvand:2017xrw} and study the spectrum of the GWs radiated during the de/confinement PT from these models for three different ranges of the bubble wall velocity. Detecting the signal of these GWs allows testing our results.\\
This paper is organized as follows: In the next section we explain properties of distinct sources for GWs generated from a first order PT. In section three, we study Hawking-Page phase transition in the AdS/QCD models and find the Gravitational wave spectrum for three different regimes of the bubble wall velocity. In the last section, we summarize the results.\\

\section{Gravitational waves of a first order phase transition}
As mentioned before, during a first order cosmological PT occurring in a thermal bath, bubbles are nucleated and because of the vacuum energy released from the initial phase, bubbles expand. In the hydrodynamical description of the bubble evolution, the bubble velocity, $ v_b $, is an important parameter which affects the GW generation of this process. Two modes of the bubble wall velocity are classified, the bubble front moving with subsonic velocity, deflagration, and supersonic velocity, detonation. For small bubble wall velocities, the big contribution for the GW energy density is not expected since the energy almost thermalize the fluid. However, for relativistic velocities, the imprint of GW sources of PTs can be traced. If the wall velocity is held at a relativistic velocity, the role of the fluid is very important and the GW contribution comes from sound waves and MHD turbulence. In the case which bubbles can run away without a bound, the energy of the runaway bubbles cannot be ignored and three sources of GWs coexist \cite{Espinosa:2010hh, Caprini:2015zlo}. In the following, we explain how to calculate the contribution of each source.\\

\subsection{Bubble collision}
After bubble nucleation and expansion, they collide with each other and the fraction of the latent heat of the system in the thermal bath is converted to GWs. The GW generated from the bubble collision is simulated by the envelope approximation \cite{Kamionkowski:1993fg} so that the anisotropic transverse component of the energy-momentum of uncollided bubble envelope, resulted from the broken spherical symmetry of a colliding bubble, is taken into account. \footnote{For an analytic approach see \cite{Jinno:2016vai}.} Numerical fits give this GW energy density as
\begin{equation}\label{spe}
h^2\Omega _{en}(f)=3.5\times 10^{-5}\Big(\frac{0.11 v_b^3}{0.42+v_b^2}\Big) \Big(\frac{H_*}{\tau}\Big)^{2}\Big(\frac{\kappa \alpha}{1+\alpha}\Big)^2 \Big(\frac{10}{g_*}\Big)^{\frac{1}{3}}S_{en}(f),
\end{equation}
where the spectral shape of the GW is \cite{Huber:2008hg}
\begin{equation}\label{s}
S_{en}(f)=\frac{3.8(\frac{f}{f_{en}})^{2.8}}{1+2.8(\frac{f}{f_{en}})^{3.8}}.
\end{equation}
The present red-shifted peak frequency is given by
\begin{equation}
f_{en}=11.3\times 10^{-9} [\mathrm{Hz}] \Big(\frac{0.62}{1.8-0.1 v_b+v_b^2}\Big)\Big(\frac{\tau}{H_*}\Big)\Big(\frac{T_*}{100~\mathrm{MeV}}\Big)\Big(\frac{g_*}{10}\Big)^{\frac{1}{6}}.
\end{equation}
The spectrum is almost a function of $ f^3 $ for small frequencies and $ f^{-1} $ for frequencies larger than the peak frequency. Also, $ \alpha $ is the vacuum energy density to the thermal energy density ratio,
\begin{equation}\label{a}
\alpha =\frac{\epsilon _*}{\frac{\pi^2}{30}g_*T_*^4}, ~~~~~~~~~~~\epsilon _*=\Big(-\Delta F(T)+T\frac{d \Delta F(T)}{dT}\Big)\Bigg|_{T=T_*}.
\end{equation}
$ \Delta F $ is the free energy difference between two phases and $ T_* $ is the temperature at which the PT takes place. Also, $ \kappa $ is the fraction of the vacuum energy  converted into the kinetic energy of the bubbles and $ \tau ^{-1} $ is the duration of the PT. Moreover, the Hubble parameter is given by
\begin{equation}\label{h}
H_*=\sqrt{\frac{8\pi^3 g_*}{90}}\frac{T_*^2}{m_{pl}},
\end{equation}
where $ g_* $ denotes the number of effective relativistic degrees of freedom, which is almost 10 at the QCD PT, and  $ m_{pl}=1.22\times 10^{22}~\mathrm{MeV} $ is the Planck mass.\\

\subsection{Sound waves and MHD turbulence}
After bubbles collided, the fraction of the energy is transformed into the plasma motion, $ \kappa _v $. This kinetic energy of the plasma generates MHD turbulence, as a Kolmogorov turbulence, which induces GW radiation. Furthermore, as proposed in \cite{Hindmarsh:2013xza}, the compression waves in the fluid, sound waves, can be another source for GW production. The GW contribution from sound waves and MHD turbulence is calculated in \cite{Hindmarsh:2015qta} and \cite{Caprini:2009yp}, respectively, as
\begin{equation}\label{sps}
h^2\Omega _{sw}(f)=5.7\times 10^{-6}\Big(\frac{H_*}{\tau}\Big)\Big(\frac{\kappa _v \alpha}{1+\alpha}\Big)^2\Big(\frac{10}{g_*}\Big)^{\frac{1}{3}} v_b~ S_{sw}(f),
\end{equation}
and
\begin{equation}\label{spt}
h^2\Omega _{tu}(f)=7.2\times 10^{-4}\Big(\frac{H_*}{\tau}\Big)\Big(\frac{\kappa _{tu} \alpha}{1+\alpha}\Big)^{\frac{3}{2}}\Big(\frac{10}{g_*}\Big)^{\frac{1}{3}} v_b~ S_{tu}(f),
\end{equation}
where $ \kappa _{tu}=\varepsilon \kappa _v $ is the fraction of the latent heat converted to MHD turbulence ($ \varepsilon $ is the fraction attributed to the turbulent fluid motion and can be of the order of 0.05 \cite{Hindmarsh:2015qta}). The spectral shapes of either source are given by \cite{Caprini:2015zlo}
\begin{eqnarray}\label{sp}
S_{sw}(f)&=&\Big(\frac{f}{f_{sw}}\Big)^3\Big(\frac{7}{4+3(\frac{f}{f_{sw}})^{2}}\Big)^{\frac{7}{2}},\nonumber \\
S_{tu}(f)&=&\frac{(\frac{f}{f_{tu}})^3}{(1+\frac{f}{f_{tu}})^{\frac{11}{3}} (1+\frac{8\pi f}{h_*})},
\end{eqnarray}
where the red-shifted Hubble frequency and peak frequency of sources are given by the following relations, respectively,
\begin{equation}
h_*=1.1\times 10^{-8} [\mathrm{Hz}]\Big(\frac{T_*}{100~\mathrm{MeV}}\Big)\Big(\frac{g_*}{10}\Big)^{\frac{1}{6}},
\end{equation}
\begin{eqnarray}
f_{sw}=1.3\times 10^{-8} [\mathrm{Hz}] \Big(\frac{1}{v_b}\Big)\Big(\frac{\tau}{H_*}\Big)\Big(\frac{T_*}{100~\mathrm{MeV}}\Big)\Big(\frac{g_*}{10}\Big)^{\frac{1}{6}},\nonumber \\
f_{tu}=1.8\times 10^{-8} [\mathrm{Hz}] \Big(\frac{1}{v_b}\Big)\Big(\frac{\tau}{H_*}\Big)\Big(\frac{T_*}{100~\mathrm{MeV}}\Big)\Big(\frac{g_*}{10}\Big)^{\frac{1}{6}}.
\end{eqnarray}
As seen from Eq. (\ref{sp}), the spectrum pertaining to sound waves and MHD turbulence is scaled approximately as $ f^3 $ with frequencies below the peak frequency for both and as $ f^{-4} $ and $  f^{-2} $ for larger frequencies, respectively.\\
Distinct regimes of the bubble wall velocity result in different contributions of the concerned GW sources \cite{Espinosa:2010hh, Caprini:2015zlo}. We study and classify the following cases:
\begin{itemize}
\item For deflagration bubbles with non-relativistic velocities, sound waves and MHD turbulence are salient sources of GWs. Therefore, $h^2\Omega (f)=h^2\Omega _{sw}+h^2\Omega _{tu} $. In this case, $ \kappa _v  $ is given by
\begin{equation}
\kappa _v =v_b^{\frac{6}{5}}\frac{6.9 \alpha}{1.36-0.037\sqrt{\alpha}+\alpha},~~~~~~v_b\ll c_s
\end{equation}
where $ c_s^2 =1/3 $.
\item  In the limit of bounded relativistic velocities, we utilize Jouguet detonations in which
\begin{equation}
\kappa _v =\frac{\sqrt{\alpha}}{0.135+\sqrt{0.98+\alpha}},~~~~~~v_b =\frac{\sqrt{\frac{2}{3}\alpha+\alpha ^2}+\sqrt{\frac{1}{3}}}{1+\alpha}.
\end{equation}
In this case also sound waves and MHD turbulence are two important sources of GWs.
\item For relativistic velocities, we can consider runaway bubbles reaching the speed of light. In this case, the energy of bubbles cannot be neglected and three GW sources should be considered, i.e $ h^2\Omega (f)=h^2\Omega _{en}+h^2\Omega _{sw}+h^2\Omega _{tu} $. The minimum value of $ \alpha $ that bubbles can run away is given by \cite{Espinosa:2010hh}
\begin{equation}\label{inf}
\alpha_{\infty}=\frac{30}{24 \pi ^2}\frac{\sum _a c_a \Delta m_a ^2}{g_* T_*^2},
\end{equation}
where $ c_a=1~(1/2)N_a $ is the number of degrees of freedom for boson (fermion) species and $ \Delta m_a $ is the mass difference of the particles between two phases.  In these bubbles, $ \alpha $ should be greater than $ \alpha _{\infty} $, and $ \kappa ,~\kappa _v $ parameters are given by
\begin{equation}
\kappa =1-\frac{\alpha _{\infty}}{\alpha},~~~~~~~~\kappa _v=\frac{\alpha _{\infty}}{0.73+0.083\sqrt{\alpha _{\infty}}+\alpha _{\infty}}.
\end{equation}
\end{itemize}

\section{AdS/QCD models}
\begin{figure}[th]
\includegraphics[scale=0.85]{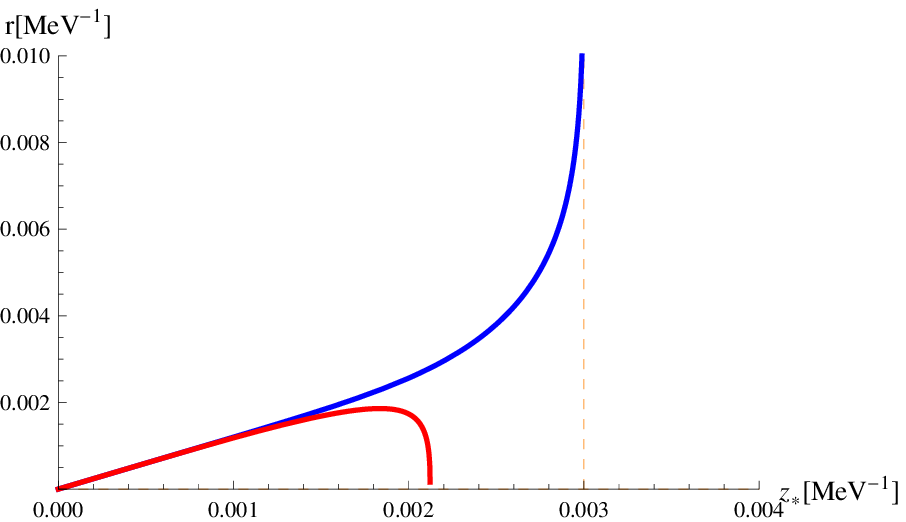} \includegraphics[scale=0.85]{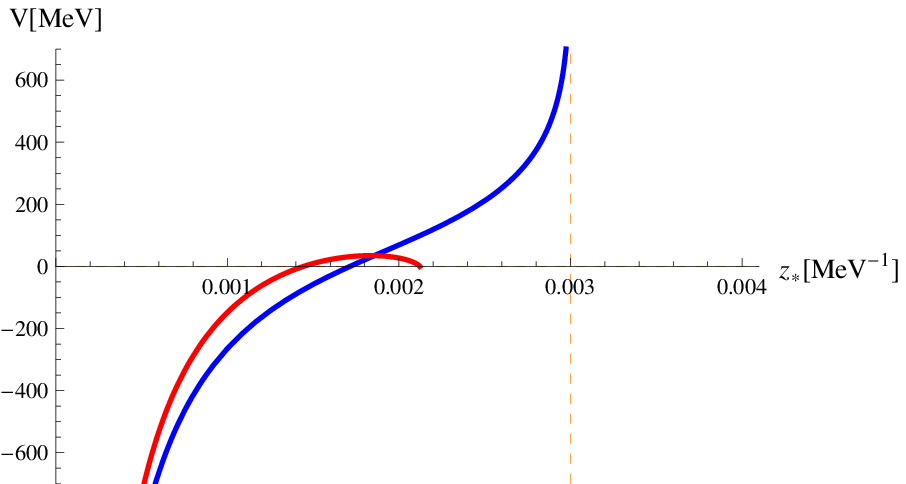}
\caption{\label{f1} We plot the quark-antiquark distance (the left plot) and potential in terms of the maximum depth of the string in the confined and deconfined phase. The blue curves which go to infinity show these quantities in the confined phase. The baryochemical potential at which the quantities tend to infinity at $ z_* $ reaching the wall is $ 500 ~\mathrm{MeV} $.}
\end{figure}

Almost $ 10^{-5}~\mathrm{secs} $ after the big bang, the quark-gluon plasma phase transformed to the color confined phase. Based on AdS/CFT correspondence, this PT corresponds to the first order H-P PT. In a medium with finite chemical potential, one can find an H-P-type PT between RN AdS BH and tc AdS space such that the baryochemical potential or quark number operator corresponds to the time component of the bulk gauge field.\\
In this section, we apply the hard wall model in which the AdS space is compactified by cutting the radial region, at $ z_0 $ corresponded to the IR cut-off in energy. The Euclidean gravitational action in five dimensions is given by
\begin{equation}\label{ac}
S=\int d^5x ~\sqrt{g}\Big[\frac{-1}{2k^2}(\mathcal{R}-2\Lambda )-\frac{1}{4g_5^2}F_{\mu\nu}F^{\mu\nu}\Big]-\frac{1}{k^2}\int d^4x ~\sqrt{\tilde{g}}\Big[\frac{1}{\sqrt{g}}\partial _{\mu}(\sqrt{g}~n^{\mu})\Big],
\end{equation}
where $ k^2=8\pi G_5 $, $ G_5 $ is the five dimensional Newton constant, $ \mathcal{R} $ is the Ricci scalar, $ \Lambda =-6/R^2 $ is the cosmological constant, $ g_5 $ is the five-dimensional gauge coupling, and $ F_{\mu \nu }=\partial _{\mu}A_{\nu}-\partial _{\nu}A_{\mu} $ is the U(1) bulk gauge field strength. The second integral is the surface action of Gibbons-Hawking, resulted from the variation principle \cite{Gibbons:1976ue}. Finally, $ n^{\mu} $ is the unit vector normal to the hypersurface and $ \tilde{g} $ is the boundary metric determinant. The Einstein-Maxwell equations of motion are obtained from Eq. (\ref{ac}) as
\begin{equation}\label{eom}
\mathcal{R}_{\mu\nu}-\frac{1}{2}g_{\mu\nu}\mathcal{R}+g_{\mu\nu}\Lambda =\frac{k^2}{g_5^2}\Big(F_{\mu\alpha}F^{\alpha}_{\nu}-\frac{1}{4}g_{\mu\nu}F_{\alpha\beta}F^{\alpha\beta}\Big),~~~~~~~~~~~ \mu =0, 1,..., 4
\end{equation}
\begin{equation}
\partial _{\mu}\Big(\sqrt{g} g^{\mu\nu}g^{\alpha\beta}F_{\nu\beta}\Big)=0.
\end{equation}
Euclidean metrics of RN AdS BH and tc AdS space as solutions of Eq. (\ref{eom}) in Poincar$\acute{\mathrm{e}}$ coordinate are given, respectively, by
\begin{equation}\label{rn}
ds^2=\frac{R^2}{z^2}\Big(f_{b, t}(z)dt^2+d\vec{x}^2+\frac{dz^2}{f_{b, t}(z)}\Big),
\end{equation}
with
\begin{equation}\label{r}
f_{b}(z)=1-\frac{z^4}{z_h^4}+q^2 z^4(z^2-z_h^2),
\end{equation}
\begin{equation}\label{tc}
f_t (z)=1+q^2 z^6,
\end{equation}
where $ z_h $ and $ q $ denote the black hole horizon radius and charge, respectively. As seen from Eq. (\ref{tc}), there is no black hole in the tc AdS space. However, the naked singularity at $ z=0 $ can be covered in the model by the wall which can also explain the confinement. We take only the time component of the bulk gauge field; hence, the solution for the Maxwell equation of motion is
\begin{equation}
A_t=i(\mu -Qz^2),
\end{equation}
where $ \mu $ is the baryochemical potential and $ Q $ is related to the black hole charge by the following relation
\begin{equation}
Q^2=\frac{3 g_5^2R^2}{2k^2}q^2.
\end{equation}
$ T_{tc}=1/\beta _{tc} $ is the tc AdS temperature and from the near horizon metric, the Hawking temperature of the black hole is
\begin{equation}\label{tem}
T_{RN}=\frac{1}{\pi z_h}\Big(1-\frac{1}{2}q^2z_h^6\Big).
\end{equation}
From Eq. (\ref{eom}), the Ricci scalar for both spaces is $ \mathcal{R}=k^2 F^2/(6g_5^2)-20/R^2 $ and due to the boundary condition at the black hole horizon on the gauge field, $ A(z_h)=0 $, $ Q $ is written as $ Q=\mu/z_h^2 $. For RN AdS BH, $ n^{\mu}=(0, 0, 0, 0, -z \sqrt{f_b}/R) $ and $ \tilde{g}=R^8 f_b/z^8 $. Therefore, the action density, denoted by $ I^{RN} $, for this space will be
\begin{eqnarray}\label{iden}
I^{RN}&=&\frac{1}{k^2}\int _0^{\beta _{RN}}dt\int _{\epsilon}^{z_h}dz~\sqrt{g}\Big(\frac{4}{R^2}+\frac{k^2}{3g_5^2}F^2\Big)-\frac{1}{k^2}\int _0^{\beta _{RN}}dt~\sqrt{\tilde{g}}\frac{1}{\sqrt{g}}\partial _{\mu}\Big(\sqrt{g}n^{\mu}\Big), \nn &=&\frac{-R^3\beta _{RN}}{k^2}\Big(\frac{3}{\epsilon ^4}-\frac{1}{z_h ^4}-\frac{2k^2}{3g_5^2R^2}\frac{\mu ^2}{z_h^2}\Big).
\end{eqnarray}
where $ \epsilon $ is the UV regulator. As it is realized from Eq. (\ref{iden}), in the $ \epsilon \rightarrow 0 $ limit, the action density is divergent. To eliminate the infinities dual to UV divergencies of the gauge theory side in these asymptotic AdS spaces, we use the holographic renormalization or the counterterm subtraction approach \cite{Emparan:1999pm} such that an extra surface integral is added to the gravitational action in order to $ I_t=I+I_{ct} $ becomes finite. The integrand of $ I_{ct} $ is constructed from $ R $, and the induced boundary metric and curvature. To cancel divergencies in Eq. (\ref{iden}), we employ the following counterterm action density:
\begin{equation}\label{ct}
I^{RN}_{ct}=\frac{1}{k^2}\int _0^{\beta _{RN}}dt ~\sqrt{\tilde{g}}\frac{3}{R}=\frac{3R^3\beta _{RN}}{k^2}\Big(\frac{1}{\epsilon ^4}-\frac{1}{2z_h ^4}-\frac{k^2}{3g_5^2R^2}\frac{\mu ^2}{z_h^2}\Big).
\end{equation}
Finally, the finite total action density is
\begin{equation}\label{tot}
I_t^{RN}=I^{RN}+I_{ct}^{RN}=\frac{-R^3\beta _{RN}}{k^2}\Big(\frac{1}{2z_h ^4}+\frac{k^2}{3g_5^2R^2}\frac{\mu ^2}{z_h^2}\Big).
\end{equation}
Thus, from Eq. (\ref{tot}) one can derive physical quantities, including the free energy density of the system, $ F^{RN}\simeq T_{RN}I^{RN}_t $. To find the PT parameters pertaining to the GW spectrum, we should proceed this approach for tc AdS space as well. To do so, we first fix $ n^{\mu}=(0, 0, 0, 0, -z\sqrt{f_t}/R) $ and $ \tilde{g}=R^8 f_t/z^8 $ for the tc AdS space, and also determine the Dirichlet boundary condition for the gauge field as $ A(z_0)=-i\mu /2 $ due to which $ \tilde{Q}=3\mu /(2z_0^2) $. Then, the action density of the space is obtained from
\begin{eqnarray}\label{ident}
I^{tc}&=&\frac{1}{k^2}\int _0^{\beta _{tc}}dt\int _{\epsilon '}^{z_0}dz~\sqrt{g}\Big(\frac{4}{R^2}+\frac{k^2}{3g_5^2}F^2\Big)-\frac{1}{k^2}\int _0^{\beta _{tc}}dt~\sqrt{\tilde{g}}\frac{1}{\sqrt{g}}\partial _{\mu}\Big(\sqrt{g}n^{\mu}\Big), \nn &=&\frac{-R^3\beta _{tc}}{k^2}\Big(\frac{3}{\epsilon '^4}+\frac{1}{z_0 ^4}+\frac{3k^2}{2g_5^2R^2}\frac{\mu ^2}{z_0^2}\Big).
\end{eqnarray}
By the counterterm action density similar to the one attained in Eq. (\ref{ct}), we remove infinities and obtain the total action density
\begin{equation}\label{ctt}
I^{tc}_{ct}=\frac{1}{k^2}\int _0^{\beta _{tc}}dt ~\sqrt{\tilde{g}}\frac{3}{R}=\frac{3R^3\beta _{tc}}{k^2}\frac{1}{\epsilon '^4},
\end{equation}
\begin{equation}\label{tott}
I_t^{tc}=I^{tc}+I_{ct}^{tc}=\frac{-R^3\beta _{tc}}{k^2}\Big(\frac{1}{z_0 ^4}+\frac{3k^2}{2g_5^2R^2}\frac{\mu ^2}{z_0^2}\Big).
\end{equation}
Setting $ \beta _{RN}=\beta _{tc} $ and $ \epsilon =\epsilon ' $, we can attain $ \Delta I $ and also $ \Delta F $ as
\begin{equation}\label{fe}
\Delta F\simeq \frac{R^3}{k^2}\Big(\frac{1}{z_0 ^4}-\frac{1}{2z_h ^4}+\frac{3k^2}{2g_5^2R^2}\frac{\mu ^2}{z_0^2}-\frac{k^2}{3g_5^2R^2}\frac{\mu ^2}{z_h^2}\Big).
\end{equation}
One may also rewrite the equation in terms of these relations: $ R^3/k^2 =N_c^2/4\pi ^2 $ and $ g_5^2= 4\pi ^2 R/(N_cN_f) $ \cite{Sin:2007ze}, where $ N_c $ and $ N_f $ are the number of colors and flavors, respectively. The H-P PT, occurring at $ \Delta F=0 $, can be found for $ z_h\leq z_0 $, where $ z_0=1/(323 ~\mathrm{MeV}) $ is obtained from the lightest $ \rho $ meson mass \cite{Polchinski:2001tt}. In AdS/QCD models, vector mesons are described by bulk vector field fluctuations, $ v_{\mu} $, in the tc AdS background in the gauge where $ v_z=0 $. These fields satisfy the following equation of motion \cite{Lee:2009bya, Park:2011qq}
\begin{equation}\label{mm}
\partial _z(\frac{f_t}{z}e ^{-\phi}\partial _z v_{\mu})+\frac{m_v^2 e^{-\phi}}{z f_t}v_{\mu}=0,
\end{equation}
where $ m_v $ is the vector meson mass and $ \phi =0 $ is the case considered in the hard wall model. The $ \mu =0 $ corresponds to $ q=0 $, and Eq. (\ref{mm}) in this case is solvable and its solutions are Bessel functions. Then, from the lightest $ \rho $ meson mass, $ z_0 $ is determined. When $\mu $ is finite, the meson spectra should be numerically studied. As \cite{Lee:2009bya} shows one can obtain that the $ \rho $ meson mass decreases, as baryochemical potential increases. One should notice that for $ \phi =c z^2 $, the case which will be studied in the next part, the equation for $ \mu =0 $ is exactly solvable but the solutions are Laguerre polynomials.\\
From Eq. (\ref{fe}), one can find a relation between baryochemical potential and the horizon radius and through this relation the PT temperature can be expressed in terms of baryochemical potential. However, in order to determine the baryochemical potential and temperature at the PT, we study the order parameter of the center symmetry of the gauge group. The relevant order parameter is the expectation value of Polyakov loop which is given by
\begin{equation}
\langle \mathcal{P} \rangle =e^{-\frac{V(T)}{T}},
\end{equation}
where $ V(T)=V(r=\infty ,T ) $ is the heavy quark and antiquark potential and $ r $ denotes their distance. In the duality context this quantity is calculated by the string world-sheet action (Nambu-Goto action), $ \langle \mathcal{P} \rangle \sim \mathrm{exp} (-S_{NG}^{on-shell}) $.\\
Two end points of the open string on the boundary of the bulk background at $ z=0 $ are considered as quark-antiquark pairs. In the confined phase, these pairs form meson states and open strings are always U-shape configuration reaching a maximum at $ z=z_* $, behind the wall. In RN AdS BH or the deconfined phase when the maximum depth of the string reaches the horizon, the string configuration becomes two straight strings corresponding to free heavy quark and antiquark.\\
The Nambu-Goto action describing an open string world-sheet is given by
\begin{equation}\label{ng}
S^{NG}=\frac{1}{2\pi \alpha '}\int d^2\xi \sqrt{\mathrm{det} g_{ab}},
\end{equation}
where $ g_{ab}=g_{\mu\nu}\partial _a X^{\mu}\partial _b X^{\nu} $ is the induced metric on the two-dimensional world-sheet and $ g_{\mu\nu} $ is the five-dimensional background metric. The coordinates $ (\xi ^0, \xi ^1) $ parametrize $ g_{ab} $ where $ (a, b) $ run over those two dimensions. We choose the static gauge: $ \xi ^0 =t $, $ \xi ^0 =x $ and $ z=z(x) $. Therefore, Eq. (\ref{ng}) will be
\begin{equation}
S^{NG}=\frac{\beta R^2}{2\pi \alpha '}\int ^{\frac{r}{2}}_{-\frac{r}{2}} dx~ \frac{\sqrt{f(z)+z'^2}}{z^2},
\end{equation}
where $ \beta $ is the inverse of temperature and depending on the background, $ f(z) $ stands for Eq. (\ref{r}) or Eq. (\ref{tc}). The quark and antiquark are located at $ (z=0, x=\mp r/2) $ and the string configuration satisfies the boundary conditions: $ z(x=0)=z_* $, $ z(x=\mp r/2)=0 $ and $ z'=dz/dx |_{x=0}=0 $. From the following relation, one can obtain the Hamiltonian of the string as a conserved quantity
\begin{equation}\label{ham}
\mathcal{H}=z'\frac{\partial \mathcal{L}}{\partial z'}-\mathcal{L}= \frac{-\beta }{2\pi \alpha '} \frac{R^2}{z^2}\frac{f(z)}{\sqrt{f(z)+z'^2}}.
\end{equation}
By using Eq. (\ref{ham}) at $ z=z_* $, $ \mathcal{H}=-\beta R^2\sqrt{f(z_*)}/(2\pi \alpha ' z_*^2)$, we find the distance between the quark-antiquark pair
\begin{equation}
r=\int ^{\frac{r}{2}}_{-\frac{r}{2}} dx=2\int _0^{\frac{r}{2}}dz \frac{1}{z'}=2\int _0 ^{z_*}dz \frac{\sqrt{f(z_*)}}{\sqrt{f(z)}}\frac{z^2}{\sqrt{f(z)z_*^4-f(z_*)z^4}}.
\end{equation}
Moreover, one can calculate the quark-antiquark potential as
\begin{equation}
V=\frac{R^2}{\pi \alpha '}\Bigg (\int _0 ^{z_*}dz \frac{\sqrt{f(z)}}{z^2\sqrt{f(z)-\frac{z^4}{z_*^4}f(z_*)}}-\int _0 ^{z_m}dz \frac{1}{z^2} \Bigg )
\end{equation}
where the second integral is added to renormalize the potential and remove divergences at $ z=0 $. The second term considered as a straight open string is attained under the condition that $ \xi ^0=t $, $ \xi ^1=z $ and $ x= \mathrm{const} $. Furthermore, $ z_m $ denotes $ z_h $ or $ z_0 $ in RN AdS BH or tcAdS background, respectively.
In Fig. \ref{f1}, we plotted the distance and potential function of the heavy quark-antiquark pair in terms of $ z_* $ in the confined and deconfined phase. In the following we explain the  results and analyze the different behavior of the string configuration based on these quantities to realize when the transition occurs.

\begin{figure}[th]
\begin{center}
\includegraphics[scale=1]{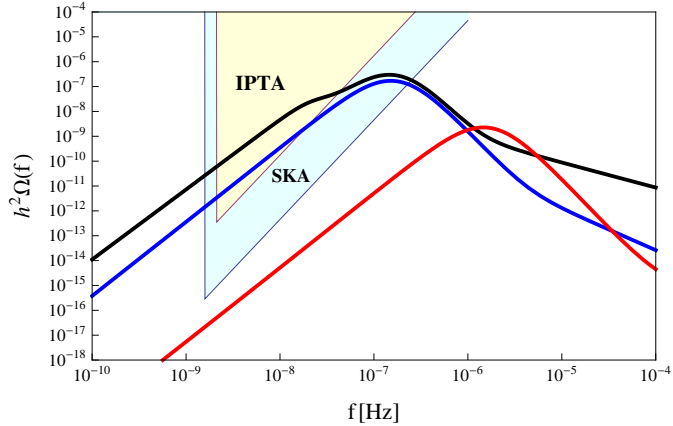}\caption{\label{f2} We display the GWs of the de/confinement PT detectable by IPTA and SKA detectors. The black curve, the top line, is the GW spectrum in the case of runaway bubbles, the blue curve denotes the GW for Jouguet detonations, and the red one, the bottom line, belongs to deflagrations with non-relativistic velocities. The sensitivity region of detectors is based on 20  pulsars with $ 10^{-7}$ secs timing precision in 15-year observation time for IPTA, and 100 pulsars with $ 3\times  10^{-8} $ secs timing precision in 20-year observation time for SKA \cite{Moore:2014lga}.}
\end{center}
\end{figure}

In the deconfined phase, the string reaches the horizon which corresponds to free quark-antiquark. During the PT or hadronization, the wall which explains the confinement appears and the string configuration changes. For $\mu = 0$, the maximum depth of the string can go beyond the wall. Thus, we can consider that the PT occurs when the string reaches the wall. Then, one can find the transition temperature through the wall, $z_0$. In the case with finite baryochemical potential, the transition temperature depends on the horizon radius and baryochemical potential, Eq. (\ref{tem}). From Eq. (\ref{fe}) horizon radius is related to baryochemical potential. Thus, the temperature is determined by $ \mu $. Taking $ \mu $ to be fixed during the PT, we need to find the baryochemical potential at which the potential goes to infinity at $ z_* $ reaching the wall. Our numerical calculations, with $ N_f =2 $ and $ N_c =3 $, show that this takes place for baryochemical potential around $ 500~ \mathrm{MeV} $. \\
This finite order of magnitude of baryochemical potential can be justified by some leptogenesis scenarios. As \cite{Schwarz:2009ii} shows with large lepton asymmetry, $ l\simeq 0.02 $, these finite baryochemical potentials are feasible although its calculations also does not include interaction effects.\\
As we see from Fig. \ref{f1}, in the confined phase $ r $ can go to infinity (blue line), while it reaches a maximum at some $ z_* $ in the deconfined phase (red line). Also, the potential can go to infinity as $ r $ tends to infinity in the confined phase and this implies that the order parameter is vanishing. On the other hand, in the deconfined phase, the potential has a maximum and the order parameter is finite. In this case, when $ z_* $ reaches the horizon the potential becomes zero. This may be interpreted as the dissociation point at which the U-shape string configuration is transformed into two straight string corresponded to two free heavy quark and antiquark \cite{Park:2009nb}.\\
Considering the baryochemical potential of the order of $ \mu =500~ \mathrm{MeV} $, one obtains from Eqs. (\ref{fe}) and (\ref{tem}) the transition temperature of the de/confinement PT, $ T_*=112 ~\mathrm{MeV} $, which is lower than the one attained in the zero chemical potential case \cite{Ahmadvand:2017xrw}.
Assuming the transition temperature is equivalent to the bubble nucleation temperature, we finally can calculate analytically the latent heat, $ \epsilon _* $, and $ \alpha $ at the transition in the presence of the baryochemical potential
\begin{equation}\label{al}
\alpha =\frac{81 N_c^5}{2(3N_c-\mu ^2N_f z_h^2)^3}.
\end{equation}
From this relation, one can figure out the PT becomes stronger in comparison with the zero chemical potential case \cite{Ahmadvand:2017xrw} and realize the effect of baryochemical potential as the source of quark number. Also, we may study how $\alpha$ changes if one considers different numbers of flavors in the model. It is found  that at fixed $ N_c $, the baryochemical potential at the transition and $ \alpha $ decrease as the number of flavors increases.\\
Moreover, in the runaway case for $ \alpha _{\infty} $, the main contribution comes from the particles that become heavy during the PT. As a result, for this PT, $ \Delta m $ can be interpreted as the quark mass difference between the constituent (effective) quark mass \cite{Lavelle:1995ty} and the quark mass in the deconfined phase, $ \Delta m\approx 400 ~\mathrm{MeV} $. Thus, with $ N_a=6 $ for quark particles, $ N_c=3 $, and $ N_f=2 $ for two heavy quarks relevant at the transition temperature, we obtain $ \alpha > \alpha _{\infty} $ which is the condition satisfied for the runaway walls \cite{Espinosa:2010hh, Chala:2016ykx} and if we assume there exist no hydrodynamic obstacles and the wall thickness is smaller than the mean free-path of the particles, these bubbles run away. However, considering other assumptions and possible bubble wall velocities, this behavior will change. Hence, for Jouguet detonations and non-relativistic velocities, we expect that $ \alpha $ decreases such that $ \alpha \lesssim \alpha _{\infty} $. For these cases we take $ \alpha $ around $ \alpha _{\infty} $ which is obtained in the model.\\
As we argued in \cite{Ahmadvand:2017xrw}, one can assume $ \tau =10 H_* $ related to the duration of the PT. Finally, putting the relevant parameters in Eqs. (\ref{spe}), (\ref{sps}), and (\ref{spt}), we can identify the GW spectrum produced during the PT.\\
In Fig. \ref{f2}, we indicate the GW generated for distinct regimes of the wall velocity. For deflagrations with $ v_b =0.1 $, the detectors will not be able to capture the signal,  while in the case of Jouguet detonations, which here reach $ v_b =0.96 $, and runaway bubbles we can expect to track down their signals in the near future.
%%%%%%%%%%%%%%%%%%%%%%%%%%%%%%%%%%%%
\subsection{The soft wall model}%%%%
\begin{figure}[th]
\begin{center}
\includegraphics[scale=1]{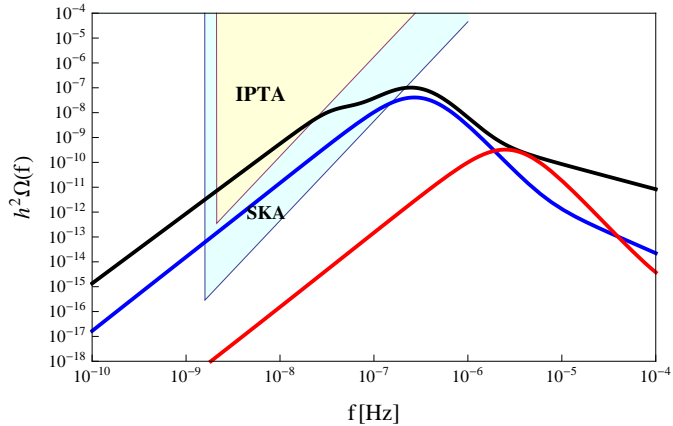}
\caption{\label{f3}The GW spectrum calculated from the soft wall model is displayed by the same conditions mentioned in Fig. \ref{f2}.}
\end{center}
\end{figure}

%%%%%%%%%%%%%%%%%%%%%%%%%%%%%%%%%%%%
In this part, we try to find the required quantities within the soft wall model in which the truncation of the space is smoothly carried out by a non-dynamical dilaton field. Therefore, We consider the same solutions, mentioned in Eq. (\ref{rn}), for the equation of motion derived from the following action. However, since imposing the IR cut-off is differently performed in the models, the results vary. The gravitational action is given by
\begin{equation}\label{acs}
S=\int d^5x ~\sqrt{g}~ e^{-\phi} \Big[\frac{-1}{2k^2}(\mathcal{R}-2\Lambda )-\frac{1}{4g_5^2}F_{\mu\nu}F^{\mu\nu}\Big]-\frac{1}{k^2}\int d^4x ~\sqrt{\tilde{g}}~ e^{-\phi} \Big[\frac{1}{\sqrt{g}}\partial _{\mu}(\sqrt{g}~n^{\mu})\Big].
\end{equation}
where the dilaton field is $ \phi =cz^2 $. The boundary condition of the gauge field at the horizon, $ n^{\mu} $ and $ \tilde{g} $ are equal to the previous model for the RN AdS BH space. Hence, the action density is obtained as follows
\begin{eqnarray}\label{idens}
I^{RN}&=&\frac{R^3}{k^2}\int _0^{\beta _{RN}}dt\int _{\epsilon}^{z_h}dz~e^{-\phi}\Big(\frac{4}{z^5}-\frac{4k^2}{3g_5^2R^2}\frac{\mu ^2z}{z_h^4}\Big)-\frac{1}{k^2}\int _0^{\beta _{RN}}dt\sqrt{\tilde{g}}~e^{-\phi} \frac{1}{\sqrt{g}}\partial _{\mu}\Big(\sqrt{g}n^{\mu}\Big) \nn &=&\frac{-R^3\beta _{RN}}{k^2}\Big[\frac{3}{\epsilon ^4}-\frac{2c}{\epsilon ^2}+c^2\ln (-c\epsilon ^2)+c^2(\frac{1}{2}+\gamma)-\frac{e^{-cz_h^2}}{z_h^4}(cz_h^2-1)-c^2\mathrm{Ei}(-cz_h^2)\nn &-&\frac{2k^2}{3g_5^2R^2}\frac{\mu ^2}{z_h^4}\Big(\frac{e^{-cz_h^2}}{c}-\frac{1}{c}+2z_h^2\Big)-\frac{2}{z_h^4} \Big].
\end{eqnarray}
where $ \mathrm{Ei}(x)\equiv -\int _{-x}^{\infty}dt~e^{-t}/t  $ and $ \gamma \sim 0.5772 $ is Euler's constant. The result has divergencies and we should supplement a counterterm action to be finite. The counterterm action density is
\begin{eqnarray}\label{cts}
I^{RN}_{ct}&=&\frac{1}{k^2R}\int _0^{\beta _{RN}}dt ~\sqrt{\tilde{g}}~e^{-\phi}\Big(3+\phi+\phi ^2\ln (-\phi)\Big)\nn &=&\frac{R^3\beta _{RN}}{k^2}\Big(\frac{3}{\epsilon ^4}-\frac{3}{2z_h ^4}-\frac{k^2}{g_5^2R^2}\frac{\mu ^2}{z_h^2}-\frac{2c}{\epsilon ^2}+c^2\ln (-c\epsilon ^2)+\frac{c^2}{2}\Big).
\end{eqnarray}
and the finite total action density will be
\begin{equation}
I^{RN}_t=\frac{R^3\beta _{RN}}{k^2}\Big[-c^2\gamma +\frac{e^{-cz_h^2}}{z_h^4}(cz_h^2-1)+c^2\mathrm{Ei}(-cz_h^2)+\frac{2k^2}{3g_5^2R^2}\frac{\mu ^2}{z_h^4}\Big(\frac{e^{-cz_h^2}}{c}-\frac{1}{c}+\frac{z_h^2}{2}\Big)+\frac{1}{2z_h^4} \Big].
\end{equation}
By the same procedure, we can also calculate the obtained quantities for the tc AdS space. Note for this space the boundary condition for the gauge field leads to $ \tilde{Q}=3c\mu /2 $, while $ n^{\mu} $ and $ \tilde{g} $ are the same as the hard wall model.
 \begin{eqnarray}
I^{tc}&=&\frac{R^3}{k^2}\int _0^{\beta _{tc}}dt\int _{\epsilon '}^{\infty}dz~e^{-\phi}\Big(\frac{4}{z^5}-\frac{3k^2c^2\mu ^2z}{g_5^2R^2}\Big)-\frac{1}{k^2}\int _0^{\beta _{tc}}dt\sqrt{\tilde{g}}~e^{-\phi} \frac{1}{\sqrt{g}}\partial _{\mu}\Big(\sqrt{g}n^{\mu}\Big) \nn &=&\frac{-R^3\beta _{tc}}{k^2}\Big(\frac{3}{\epsilon  '^{4}}-\frac{2c}{\epsilon '^{2}}+c^2\ln {(-c\epsilon '^{2})}+c^2(\frac{1}{2}+\gamma)+\frac{2k^2\mu ^2c}{3g_5^2R^2} \Big)
\end{eqnarray}
The counterterm action density and $ I_t^{tc} $ are obtained as
\begin{eqnarray}
I^{tc}_{ct}&=&\frac{1}{k^2R}\int _0^{\beta _{tc}}dt ~\sqrt{\tilde{g}}~e^{-\phi}\Big(3+\phi+\phi ^2\ln (-\phi)\Big)=\frac{R^3\beta _{tc}}{k^2}\Big(\frac{3}{\epsilon  '^{4}}-\frac{2c}{\epsilon '^{2}}+c^2\ln {(-c\epsilon '^{2})}+\frac{c^2}{2}\Big) \nn I_t^{tc}&=&\frac{R^3\beta _{tc}}{k^2}\Big(-c^2\gamma -\frac{3k^2c\mu ^2}{2g_5^2R^2}\Big)
\end{eqnarray}
Again by setting $ \beta _{RN}=\beta _{tc} $ and $ \epsilon =\epsilon ' $, we can read off $ \Delta F $
\begin{equation}\label{fes}
\Delta F\simeq \frac{R^3}{k^2}\Big[\frac{e^{-cz_h^2}}{z_h^4}(cz_h^2-1)+c^2\mathrm{Ei}(-cz_h^2)+\frac{2k^2}{3g_5^2R^2}\frac{\mu ^2}{z_h^4}\Big(\frac{e^{-cz_h^2}}{c}-\frac{1}{c}+\frac{z_h^2}{2}\Big)+\frac{1}{2z_h^4}+\frac{3k^2c\mu ^2}{2g_5^2R^2}\Big].
\end{equation}
The H-P PT is realized from $ \Delta F=0 $, and from this equation $ z_h $ is related to $ \mu $.\\
As mentioned in the previous section, the meson mass can be attained from Eq. (\ref{mm}). Here, $ \phi =cz^2  $ and for $ \mu =0 $ solutions are Laguerre polynomials. From the lightest $ \rho $ meson mass, $ \sqrt{c}=388~ \mathrm{MeV} $ \cite{Karch:2006pv}. In the case of finite baryochemical potential, one can show from the numerical calculation meson mass increases when $ \mu $ is increased \cite{Park:2011qq}.\\
Moreover, similar to the previous model arguments, to find the temperature and baryochemical potential at the PT, we study the behavior of the order parameter during the PT. Note that in the soft wall model in order to figure out the heavy quark-antiquark potential, we should apply the positive warp factor for the model. This means the Nambu-Goto action is given by
\begin{equation}
S^{NG}=\frac{1}{2\pi \alpha '}\int d^2\xi ~e^{cz^2}\sqrt{\mathrm{det} g_{ab}}.
\end{equation}
By following the same procedure and discussion mentioned for the hard wall model, we find the baryochemical potential at the transition is around $ 100 ~\mathrm{MeV} $.
 From $ \sqrt{c}=388 ~\mathrm{MeV} $ and $ \mu =100 ~\mathrm{MeV} $, the transition temperature becomes $ T_* =192 ~\mathrm{MeV} $.\\
 Also, from Eq. (\ref{fes}) we can numerically compute the latent heat and $ \alpha $. (Here, also the phase transition would be stronger than the zero chemical potential case.) Different IR cut-off in the models gives rise to different values for the parameters, including the latent heat, due to which distinct GW spectra are attained.\\
 We can also calculate $ \alpha _{\infty} $ and for Jouguet detonation and non-relativistic velocity cases we expect $ \alpha $ reach around this $ \alpha _{\infty} $.  
As seen from Fig. \ref{f3}, for $ \tau =10H_* $, the PT GWs detectable by IPTA and SKA are in Jouguet detonation, reaching $ v_b =0.92 $, and runaway regimes whose peak freaquencies are around $ 10^{-7}~\mathrm{Hz} $.

\section{Summary}
We studied the cosmic de/confinement PT in a dense matter medium through hard and soft wall models of AdS/QCD. Finite baryochemical potential at the cosmic QCD PT scale can be allowed in the leptogenesis scenarios with the large lepton asymmetry.\\
While with the finite baryochemical potential lattice QCD would confront the sign problem, we here used AdS/QCD and considered RN AdS BH dual to the deconfinement phase and tc AdS space dual to the hadronic phase in the gauge theory side. We employed the holographic renormalization method to obtain the free energy density of each phase, within the hard and soft wall models, such that the divergencies are removed by appropriate counterterms and one can gain other thermodynamical quantities of the phases from these energy densities. We found the H-P PT, corresponding to the de/confinement PT. To determine the temperature and baryochemical potential at the transition and to realize when the PT takes place, we also studied the expectation value of Polyakov loop as the order parameter. By combining these structures, we found a better picture of the PT to obtain required quantities during the PT. Moreover, the transition temperature and latent heat at the PT were calculated analytically and numerically for the hard and soft wall models. In comparison with the zero chemical potential case, it is found that the PT is stronger. \\
We studied GW spectra in three different regimes of the bubble wall velocity and examine the reliability of consequent calculations by gravitational wave experiments. We described the contribution of the GW sources during the PT, bubble collisions, sound waves, and MHD turbulence and determined GW spectra in the models for possible bubble wall velocities. We obtained $ \alpha > \alpha _{\infty} $ which is the criterion for the runaway bubbles and if we assumed there are no hydrodynamic obstructions, bubbles could accelerate without a bound. Thus, considering other assumptions, we attained the GW for a terminal relativistic wall velocity by taking Jouguet detonations and for deflagrations with a non-relativistic velocity. To provide an ability to prove or rule out the suggested scenarios, it was indicated that IPTA and SKA detectors, using pulsar arrays, will be able to detect the signal of the GWs in the case of Jouguet detonation and runaway modes, whereas for the case of deflagration the signal cannot be captured.
\newpage

\textbf{Acknowledgments}\\

KBF acknowledges the University of Southampton for their hospitality during the course of this work.

%%%%%%%%%%%%%%%%%%%%%%%%%%%%%%%%%%
\end{document}